\documentclass[fleqn,12 pt]{article}
\usepackage{amssymb,amsmath,euscript}
\begin{document}
\newcommand{\E}{\mathcal{E}}
\newcommand{\J}{\mathbb{J}}
\newcommand{\s}{\mathcal{S}}
\newcommand{\T}{\mathbb{T}}
\newcommand{\B}{\mathbb{S}}
\newcommand{\D}{\partial}
\title{\normalsize\textbf{ Equations of one-dimensional hydrodynamics
with quantum thermal fluctuations taken into account}}
\author{\small\textbf{ O.N. Golubjeva$^1$, A.D.  Sukhanov$^2$, and V.G. Bar'yakhtar$^3$}}
\date {}
\maketitle
\begin {abstract}
We propose a generalization of equations of quantum mechanics in the
hydrodynamic form by introducing the terms taking into account the
diffusion velocity at zero and finite temperatures and the density
energy of diffusion pressure  of the thermal vacuum into the
Lagrangian density. Based on this, for a model of one-dimensional
hydrodynamics, we construct a system of equations that are similar
to the Euler equations but taking quantum and thermal effects into
account. They are a generalization of equations of Nelson's
stochastic mechanics and can be used to describe a new matter state,
namely, nearly perfect fluidity.

\textbf{Key words }: $(\hbar,k)$~dynamics, quantum thermostat,
cold vacuum, thermal vacuum, effective action, self-diffusion,
drift and diffusion velocities

\begin{center}
$^1$People's Friendship University of Russia, Moscow, Russia.
E-mail: ogol@oldi.ru\\ $^2$Bogoliubov Theoretical Physics
Laboratory, Joint Institute for Nuclear Research, Dubna, Russia.
E-mail: ogol@mail.ru\\$^3$ Institute of Magnetism, National Academy
of Sciences of Ukraine, Kiev, Ukraine. E-mail:
\end{center} \normalfont
\end {abstract}

\section*{ \small 1. Introduction }
The explanation of the universal property of matter, namely,
nearly perfect fluidity (NPF) [1], [2], discovered experimentally
in different media, such as quark--gluon plasma, ultracold gases
in traps, liquid helium, and even graphene, have attracted the
attention of many researchers in the last several years [3].
Particular theoretical approaches were used for this purpose.
However, there is no explicit basis combining them in the published
papers. Only one fact makes them close to each other, namely,
the presence of the combination of the Planck and Boltzmann
world constants in the form~$ \varkappa=\hbar/2k_B$ in the
final result compared with the experimental data.

Irrespective of the NPF study, it was established that this quantity
appears in the description of many quantum thermal phenomena,
including transport processes. In addition, the characteristic of
the effective environmental ("thermal"\; vacuum) influence in the
equilibrium state is expressed in terms of~$\varkappa$ in $
(\hbar,k)$~dynamics, which is a modification of quantum theory at
finite temperatures [4], [5].

Because the case in point is the description of the entire class of
phenomena in which the nonadditivity of influences of the quantum
and thermal types is manifested, the theory to be developed must be
independent of particular concepts of medium structure as much as
possible, which is typical of thermodynamics. In particular, to
solve the NPF problem, hydrodynamics as a part of nonequilibrium
thermodynamics must be generalized, primarily, to the quantum
region. Although fluctuations in hydrodynamics have been taken into
account during the last fifty years [6], at present, there is no
consistent theory that takes quantum and thermal effects into
account simultaneously.

In this paper, we propose to construct the holistic stochastic
hydrodynamics in a new way by starting from the microdescription.
To do this, we proceed from the hydrodynamic form of quantum mechanics,
which is well known at zero temperature. We generalize it naturally to
the case in which self-diffusion in the cold and thermal vacua are
taken into account explicitly. For the first time, this allows extending
the hydrodynamic form of quantum mechanics to finite temperatures and
taking into account not only self-diffusion manifested at zero temperature too,
but also the energy density of the diffusion pressure of the thermal vacuum.

As a result, for a one-dimensional model, we obtained a system of
equations of stochastic hydrodynamics that is applicable at any
temperature. Its distinction from the standard theory is that
it takes quantum and thermal fluctuations into account nonadditively.
Moreover, we managed to write these equations in the form of equations
of two-velocity hydrodynamics, which is a generalization of Nelson's stochastic mechanics.

\section*{\small 2. Effective action as a universal characteristic of transport processes}

We recall that in the framework of~$ (\hbar,k) $~dynamics,
we introduced a new macroparameter, namely, the effective action of the quantum
thermostat  on a system~[5]
\begin{equation}\label{1}
\J=\frac{\hbar}{2}\sqrt{\alpha^2+ 1}=\J^0\Upsilon,
\end{equation}
where
$$\alpha^2\equiv\sinh^{-2}\varkappa\frac{\omega}{T};
\;\;\;\Upsilon= \coth(\varkappa\frac\omega T),$$  and $\J^0=\hbar/2$
is the limiting values of~ $\J$ at the Kelvin temperature~ $T
\rightarrow 0.$

The most important macroparameters (the effective temperature,
the effective internal energy, and the effective entropy) are
expressed in terms of this quantity in the equilibrium case:
\begin{equation}\label{2}
 \T=\frac {\omega}{k_B}\J;
\end{equation}
\begin{equation}\label{3}
\mathbb U=\omega \J;
\end{equation}
\begin{equation}\label{4}
\B=\B^0\{1+\ln\frac{\J}{\J^0}\},
\end{equation}
where $\B^0=k_B$ is the limiting value of~ $\B$ as~$T
\rightarrow 0.$

In this connection, it would be natural to assume the
limiting value of the ratio of two fundamental quantities,
namely, the effective action~ $ \J $ and the effective entropy~ $В,
\B $
\begin{equation}\label{11=5}
\varkappa=\frac\hbar{2k_B}\equiv \left.\lim
\right|_{\scriptscriptstyle {T\rightarrow 0}}\;\frac{\J}{\B}
\end{equation}
to be the physical definition of the universal constant~$
\varkappa $.

However, this is not all. As shown below, effective transport
coefficients, which are typical of nonequilibrium thermodynamics,
can also be expressed in terms of this quantity. The latter can be
seen if the self-diffusion process occurring in a medium with the
inhomogeneous density after the temperature equilibrium is
established is used as an example.

Indeed, it was shown in the theory of Brownian motion at
sufficiently high temperatures~ [7] that the uncertainties relation
of the form
\begin{equation}\label{5=6}
(\Delta p)\;(\Delta q)
    =  mD_T
\end{equation}
is valid in this case (for~$t\gg\tau$).
Here, $D_T$ is the coefficient of purely thermal diffusion;
in this case, $D_T=k_BT\tau/m$ for a free microparticle ($\tau$
is the relaxation time) and $D_T=k_BT/m\omega$ for a Brownian oscillator~[8].

As shown in~ [9], the Schr\"odinger uncertainties relation for the
quantum oscillator in the state of equilibrium with the thermal
vacuum has the form
\begin{equation}\label{7}
(\Delta p)\;(\Delta q)=\J=
     (\frac\hbar 2 \Upsilon).
\end{equation}

Comparing~ (6) and~(7), we write this relation in the form
\begin{equation}\label{7=8}
(\Delta p)\;(\Delta q)=mD_{ef}.
\end{equation}
Then it is natural to call the quantity
\begin{equation}\label{8=9}
D_{ef}= D_{qu}\Upsilon\equiv\frac{\J}{m}
\end{equation}
the effective sefl-diffusion coefficient or the effective
action on the mass unit and the quantity~$D_{qu}=\hbar/2m$
the quantum diffusion coefficient in the cold vacuum.
The quantity~ $D_{qu}$ was introduced previously,
in particular, by Nelson~[10].

Obviously, the coefficient~ $ D_{ef}   $ acquires the physical
meaning of the effective action per mass unit. The limiting
values of~ $D_{ef}$ at high  ($k_BT\gg\hbar\omega/2$) and low ($
k_BT\ll\hbar\omega/2$) temperatures are equal to
\begin{eqnarray}\label{10}
D_{ef}\rightarrow D_T=k_BT/m\omega;\;\;\;  D_{ef}\rightarrow
D_{qu}\equiv\frac\hbar {2m},
\end{eqnarray}
respectively.

Proceeding from relation~ (8), all other effective transport
coefficients can be expressed in terms of~ $\J$. We recall that
according to the kinetic theory, the diffusion coefficient~ $D_T$
has the meaning of the kinematic viscosity coefficient so that for
shear viscosity, the formula
$$ \eta=D_T\rho_m,$$ where $   \rho_m $ is the bulk mass density, is valid.

Assuming that the interrelation between the shear viscosity
coefficient~$\eta_{ef}$  and the effective self-diffusion
coefficient~$D_{ef}$ of form (9) has a similar form, we obtain

\begin{equation}\label{11}
\eta_{ef}=D_{\scriptscriptstyle{ef}}\;\rho_{m}=\frac {\J}{
V},\;\;\;\frac1V\mathbb C_VD_{ef}.
\end{equation}
 Thus, $\eta_{ef}$  has the physical meaning of the specific
 effective action.

In our opinion, the ratio of the heat conductivity
coefficient to the conduction coefficient in the Wideman--Franz law
\begin{equation}\label{12}
  \frac{\lambda}{\sigma}= \gamma (\frac{k_B}{e})^2\,T=\gamma
  \frac{k_B}{e^2}(k_BT),
\end{equation}
where $\gamma$ is a numerical coefficient, is also
interesting. It is obvious that the presence of the
factor~$k_BT$ in this formula means that the classical
thermostat model was initially assumed in this law.

In our opinion, the generalization of this law to the
quantum thermostat model must have the form
\begin{eqnarray}\label{13}
 \frac{\lambda_{ef}}{\sigma_{ef}}= \gamma\frac{k_B}{e^2}
 (k_B\mathbb T) =
\gamma\frac{k_B}{e^2}   \omega \J= \gamma\frac{k_B}{e^2}
\omega mD_{ef}  .
\end{eqnarray}
Thus, the majority of transport coefficients can be
expressed in terms of the effective self-diffusion
coefficient~$D_{ef},$ which can, in principle, be obtained experimentally.

As for the constant~$\varkappa,$ in the analysis of
particular experiments, it can be expressed in terms
of observed transport coefficients by relations of the type

\begin{equation}\label{14}
\varkappa=\left(\frac{D_{ef}}{\B/m}\right)
_{min}=\left(\frac{\eta_{ef}}{\B/V}\right)_{min}= ...,
\end{equation}
where $\B/m$ is the effective entropy of the mass unit
and $\B/V$ is the effective entropy of the volume unit.

\section*{\small 3.    Standard quantum mechanics in the hydrodynamic form}
The standard quantum mechanics in the nonrelativistic field form (at~$ T= 0 $) can be obtained if the variation of the action functional
\begin{equation}\label{15}
\EuScript{S}=\int^{t_2}_{t_1}\;dt\int dq\;\mathcal
L_{\scriptscriptstyle{0}} [\psi^*;\psi]
\end{equation}
becomes zero [11]. Here, $\mathcal L_{\scriptscriptstyle{0}}
[\psi^*;\psi] $ is the Lagrangian density for one spinless
particle at~$T=0$, and
$\psi(q,t) $ and~$\psi^*(q,t) $ are the wave function and
its complex conjugate function, which have the meaning of
independent nonrelativistic fields. (We restrict ourselves
to the one-dimensional case.)

It is obvious that the functional $\mathcal L_{\scriptscriptstyle{0}}
[\psi^*;\psi]$ in the general case must be chosen in the form

\begin{equation}\label{16}
\mathcal L_{\scriptscriptstyle{0}}[\psi^*;\psi]=\psi^*
(q,t)\left(i\hbar\frac{\partial}{\partial t}
+\frac{\hbar^2}{2m}\;\frac{\partial^2}{\partial q^2}\right)
\psi(q,t)-\psi^* (q,t)U(q)\psi (q,t),
\end{equation}
where the nonrelativistic limit of the Klein--Gordon operator is in
the parentheses on the right, and the potential energy operator $
U(q)    $ characterizes the regular influence energy.

Independently varying the action of form (15) with
respect to the field~$\psi^* $ leads to the condition
\begin{equation}\label{17}
\int^{t_2}_{t_1}\;dt\int dq\frac{\delta\mathcal
L_{\scriptscriptstyle{0}}[\psi^*;\psi]} {\delta
\psi}=\int^{t_2}_{t_1}\;dt\int
dq\left(i\hbar\frac{\partial\psi}{\partial t}
+\frac{\hbar^2}{2m}\;\frac{\partial^2\psi}{\partial
q^2}-U(q)\psi\right)=0,
\end{equation}
which leads to the Schr\"odinger equation
\begin{equation}\label{18}
i\hbar\frac{\partial\psi}{\partial
t}=-\frac{\hbar^2}{2m}\;\frac{\partial^2 \psi}{\partial
q^2}+U(q)\psi.
\end{equation}
Accordingly, its complex conjugate equation is obtained
by varying the action of form~(15)     with respect to~$ \psi
$ and differs from formula~ (18) by the replacement of~$   i $
with~$  -i $ and~$ \psi $ with~$ \psi^*.  $ We stress that the
Schr\"odinger equations for the complex wave functions~ $  \psi $ and~$
\psi^* $ have the meaning of the Lagrange--Euler equations;
in this case, the wave functions are always complex in the
full-scale quantum mechanics.

We now represent the wave function in the form

\begin{equation}\label{19}
\psi(q,t)=\sqrt{\rho(q,t)}\exp\{i\theta(q,t)\},
\end{equation}
where $\rho (q,t)=|\psi(q,t)|^2.$ This expression, together
with its complex conjugate expression, can be substituted
directly in the Schr\"odinger equations for~ $ \psi $ and~$
\psi^* $, and the system of equations for the functions~
$\rho(q,t) $ and~ $\theta(q,t) $, which is well known in
published papers as quantum mechanics in the hydrodynamic
form, can be obtained [11],[12].

Because our aim is to construct modified hydrodynamics based
on the microdescription, we propose another approach to the
problem. It requires developing the theory in the Lagrange
formulation from the very beginning. Therefore, we start from
the transformation of the Lagrangian density~ $\mathcal
L_{\scriptscriptstyle{0}}$ by passing to variables that are
more adapted to the hydrodynamic description. As functional
arguments of the Lagrangian density, we chose two independent
real functions, namely, the probability density~$\rho$ and the
phase~ $ \theta $ instead of the complex wave functions~$ \psi
$ and~$ \psi^* $. They are similar to the functions of the
mass density~ $ \rho_m $ and the drift velocity~ $ v
\sim\dfrac{\partial \theta}{\partial q} $, which are typical
of the standard hydrodynamics.

To do this, we replace the arguments in Lagrangian density~
(16) by substituting expression~(19) and the corresponding
expression for~ $  \psi^* $ in it. After the substitution,
we obtain
\begin{eqnarray}\label{20}
\mathcal L_{\scriptscriptstyle{0}}[\psi;\psi^*]=\mathcal
L_{\scriptscriptstyle{0}}[\rho;\theta]=-\hbar\frac{\partial\theta}{\partial
t}\rho\;-\frac{\hbar^2}{2m}\left(\frac{\partial
\theta}{\partial
q}\right)^2\rho-\frac{\hbar^2}{8m}\left(\frac{\partial
\rho}{\partial
q}\right)^2\frac{1}{\rho}-\nonumber\\-U(q)\rho+i\frac{\hbar}{2}\frac{\partial\rho}{\partial
t}+\frac{\hbar^2}{2m}\cdot\frac{\partial}{\partial q}\left(
\frac 12 \frac{\partial \rho}{\partial
q}+i\rho\;\frac{\partial\theta}{\partial q}\right)
\end{eqnarray}
Here, the term containing~ $\frac{\partial\rho}{\partial t}$
can be neglected because its contribution is zero after
variation of the action~$\EuScript{S}$ of form(15) with
respect to~ $ \theta $ and~$ \rho$. The last term in~ (20) is
the total derivative with respect to~ $q,$ so that it can also
be excluded from the definition of~ $\mathcal
L_{\scriptscriptstyle{0}}[\rho;\theta]$. Therefore, as the
expression for the Lagrangian density~$\mathcal
L_{\scriptscriptstyle{0}}[\rho;\theta] $, we finally assume
\begin{equation}\label{21}
\mathcal
L_{\scriptscriptstyle{0}}[\rho;\theta]=-\hbar\;\frac{\partial\theta}{\partial
t}\;\rho\; -\frac{\hbar^2}{2m}\left(\frac{\partial \theta}{\partial
q}\right)^2 \rho\;-\;\frac{\hbar^2}{8m}\left(\frac{\partial
\rho}{\partial q}\right)^2\frac{1}{\rho}\;-U(q)\rho.
\end{equation}

Varying the action~$\EuScript{S}$ of form~(15) , in which
$\mathcal L_{\scriptscriptstyle{0}}[\rho;\theta] $ now
has form~(21), with respect to the variables~ $ \theta $
and~$\rho $ successively, we obtain the equations for
the real functions~$\rho(q,t)$ and~$\theta(q,t) $:
\begin{equation}\label{22}
\frac{\partial\rho}{\partial t}+\frac{\partial}{\partial
q}\left(\rho\frac{\hbar}{m}\;\frac{\D\theta}{\D q}\right)=0,
\end{equation}
\begin{equation}\label{23}
\hbar\frac{\partial\theta}{\partial
t}+\frac{\hbar^2}{2m}\left(\frac{\partial \theta}{\partial
q}\right)^2+U(q)-\frac{\hbar^2}{8m}\left[\frac{1}{\rho^2}\left(
\frac{\partial\rho}{\partial q} \right)^2+2\frac{\partial}{\partial
q}\left(\frac{1}{\rho}\;\frac{\partial \rho}{\partial
q}\right)\right]=0.
\end{equation}

These equations coincide with the equations that could
be obtained for the functions~ $ \rho$      and~$  \theta $
directly from the Schr\"odinger equations. However, it is
now clear that they have the meaning of the Lagrange--Euler
equations for the action~$\EuScript{S}$ of form~(15)
expressed in terms of the variables~$\rho$ и $\theta$.

It is assumed traditionally that Eq.~ (22) is the continuity
equation for~ $\rho (q,t) $. In its turn, Eq.~(23) taking
into account that the quantity~ $ \hbar\theta(q,t) $ has the
dimension of action is an analogue of the Hamilton--Jacobi
equation. In this case, the term in brackets in formula~(23)
is interpreted sometimes as an additional energy~$U_{qu}(q)$
of quantum nature vanishing in the semiclassical limit as~ $\hbar \rightarrow 0 $.

Of course, Eqs.~ (22) and~(23) for~$ \rho $ and~$ \theta $ and the
Schr\"odinger equations for~$ \psi $ and~$\psi^*$ are equivalent
formally. However, the derivation of equations of quantum mechanics
in hydrodynamic form~ (22) and~(23) directly from the principle of
least action is physically more preferable in the construction of
stochastic hydrodynamics. At the same time, to obtain the sought
result, the problem of the form of the Lagrandian density must be
solved. In our opinion, it must take the stochastic environmental
influence (quantum thermostat) into account consistently.

\section*{\small 4. Quantum self-diffusion in the "cold" vacuum}

To reveal the possibilities of generalizing~ $
\mathcal L_{\scriptscriptstyle{0}}[\rho;\theta]   $,
we first consider the case of the cold vacuum. To do this,
we make the second and third terms in expression~(21) have
the physical meaning. According to the terminology
introduced by Kolmogorov~[13] for the Markov processes
in the general theory of stochastic processes and used
by Nelson~[10] in his stochastic mechanics, we call the quantity
\begin{equation}\label{24}
v\equiv\frac{\hbar}{m}\;\frac{\D\theta}{\D q}
\end{equation}
the drift velocity. Accordingly, we call the quantity
\begin{equation}\label{25}
u\equiv-D_{qu}\frac 1\rho\;\frac{\D\rho}{\D
q}=-\frac{\hbar}{2m}\;\frac{1}{\rho}\;\frac{\D\rho}{\D q}
\end{equation}
the diffusion velocity in the cold vacuum.

If the velocities~ $v$ and~$ u $ are used, formulas~(21)-(23)
can be written in the forms
$$\mathcal L_{\scriptscriptstyle{0}}[\rho,\theta]
=-\hbar\;\frac{\D\theta} {\D t}\;\rho-\frac
m2(v^2+u^2)\rho-U\rho
;\;\;\;\;\;\;\;\;\;\;\;\;\;\;\;\;\;\;\;\;\;(21a)$$

$$\frac{\D\rho}{\D t}\;+\frac{\D}{\D q}\;(\rho v)=0;\;\;\;\;\;\;\;\;\;\;\;\;\;\;\;\;\;\; (22a)$$

$$\hbar\frac{\D\theta}{\D t}+\frac m2\;v^2+U-\frac m2[u^2-\frac{\hbar}{m}\frac{\D u}{\D
q}]=0,\;\;\;\;\;\;\;\;\;\;\;\;\;\;\;\;\;\;\;\;\;\;(23a)$$
which open opportunities for generalization.

It follows from formula~ (22а) that standard continuity
equation~(22) is quasiclassical because the probability flux
density in it depends only on the drift velocity~$  v $, while
the diffusion velocity~$ u $ produced by the stochastic action
of the cold vacuum is not taken into account in it.

In this connection, we recall that the Fokker--Planck equation~[14]
\begin{equation}\label{26}
\frac{\D\rho}{\D t}+\frac{\D}{\D q}(\rho V)=0,
\end{equation}
which contains the total velocity of the probability flux density
\begin{equation}\label{27}
V=v+u,
\end{equation}
is a more general continuity equation according to
Kolmogorov~[13]. We show that it allows describing
the approach to the thermal equilibrium state
because of self-diffusion in the cold vacuum too.

We call attention to the fact that the
combination~$\dfrac m2(v^2+u^2)$ in expression~ (21а)
for~$\mathcal L_{\scriptscriptstyle{0}}[\rho,\theta] $
is the sum of independent contributions of the kinetic
energies of the drift and diffusion motions. At the same time,
the probability flux depends on the total velocity of form~ (27).
In connection with this, to obtain the Fokker--Planck equation
in expression~(21а), the natural replacement of~$ (v^2+u^2)
$ with~$V^2$ suggests itself, which allows taking into account
the total expression for the kinetic energy related to the
probability flux. Thus, even the standard quantum mechanics
(at~$T=0$) admits the opportunities for generalization.

Thus, we generalize the Lagrangian density~ $\mathcal
L_{\scriptscriptstyle{0}}[\rho,\theta]$ of form~(21а) using the corresponding replacement. Then we obtain
\begin{eqnarray}\label{28}
\tilde{\mathcal
L}_{\scriptscriptstyle{0}}[\rho,\theta]=-\hbar\;\frac{\D\theta}{\D
t}\;\rho-\frac m2V^2\rho-U\rho=\nonumber\\=\mathcal
{L}_{\scriptscriptstyle{0}}[\rho;\theta]-mvu\rho=\mathcal
{L}_{\scriptscriptstyle{0}}[\rho;\theta]+\frac{\hbar^2}{2m}\;\frac{\D\theta}{\D
q}\;\frac{\D\rho}{\D q}.
\end{eqnarray}

Varying the action functional~$\EuScript{S}$ of form (15)
with~$\tilde{\mathcal L}_{\scriptscriptstyle{0}}[\rho,\theta]
$ with respect to по~$ \theta$ automatically leads to the
Fokker--Planck equation with the quantum diffusion coefficient
\begin{equation}\label{29}
\frac{\D\rho}{\D t}+\frac{\D}{\D q}(\rho V)=\frac{\D\rho}{\D
t}+\frac{\D}{\D q}\left(\rho\;\frac{\hbar}{m}\;\frac{\D\theta}{\D
q}\right)-D_{qu}\frac{\D^2\rho}{\D q^2}=0.
\end{equation}
At the same time, varying~$\EuScript{S}$ with respect to~$  \rho $
remains the Hamilton--Jacobi equation almost unchanged; one insignificant term
\begin{equation}\label{30}
\hbar\frac{\D\theta}{\D t}+\frac m2v^2+U-\frac
m2\left(u^2-\frac\hbar m \frac{\D u}{\D
q}\right)+\frac{\hbar}{2}\;\frac{\D v}{\D q}=0
\end{equation}
appears in the Hamilton--Jacobi equation compared with~(23a).

Obtained system of equations~(29) and~(30) generalizes Eqs.~(22а)
and~(23а), allowing one to take the quantum stochastic influence of
the cold vacuum into account consistently.

\section*{\small 5. Self-diffusion in the quantum thermostat for~$  T\ne0$}
We now use our developed approach to the description of
self-diffusion with the simultaneous inclusion of the quantum and
thermal effects. For this purpose, we introduce the Lagrangian
density~$\tilde{\mathcal L}_{\scriptscriptstyle{T}}[\rho,\theta] $,
which is temperature-dependent, and demand that it transform into
the expression~$\tilde{\mathcal
L}_{\scriptscriptstyle{0}}[\rho,\theta] $ of form~(28)
as~$T\rightarrow 0 $. To do this, it suffices to replace the
diffusion coefficient~ $D_{qu}$ with~$ D_{ef}$ of form~(8) in
expression~(25) for the diffusion velocity and to introduce, in the
expression for the Lagrangian density, the additional
term~$U_T(q)\rho$ taking into account the energy density of the
diffusion pressure because of the thermal environmental stochastic
influence.

According to our reasoning, the expression for~$U_T$ must have a
form that is similar to that of the factor~$- mu^2/2$ in the cold
vacuum~(21а). However, it must be modified so that
$ U_T\rightarrow 0$ as~$ T\rightarrow 0$. We introduce it as follows:
\begin{equation}\label{30 (31)}
U_T(q)=-\frac
m2\left[\frac{\alpha}{\Upsilon}\right]^2u_{ef}^2=-
\frac{\hbar^2}{8m}\;\alpha^{2}
\left(\frac{1}{\rho}\;\frac{\D\rho}{\D q}\right)^2,
\end{equation}

$$\mbox{where}\;\;\alpha^2\equiv\sinh^{-2}\varkappa\frac{\omega}{T};
\;\;\;\Upsilon\equiv \coth(\varkappa\frac\omega T),$$
$$\mbox{and also }\;\;\;u_{ef}\equiv -D_{ef}\dfrac{1}{\rho}\dfrac{\D\rho}{\D q}$$ is
the effective diffusion velocity in the thermal vacuum
with~$D_{ef}$ of form~(9), which is determined in analogy with
the velocity~ $u$ in~(25).

Thus, as the Lagrangian density for~ $ T\ne 0$, we chose the expression
\begin{equation}\label{32} \tilde{\mathcal
L}_T(\rho,\theta)=-\hbar\frac{\D\theta}{\D t}\;\rho -\frac
m2(v+u_{ef})^2\rho-U\rho-U_{\scriptscriptstyle T}\rho.
\end{equation}
For the convenience of further variation, we rewrite
expression~(32) in the explicit form in terms of the
random functions~$\theta$  and~$\rho$:
\begin{multline}\label{33}
\tilde{\mathcal L}_T(\rho,\theta)=-\hbar\;\frac{\D\theta}{\D
t}\;\rho -\left\{\frac{\hbar^2}{2m}\;\left(\frac{\D\theta}{\D
q}\right)^2\rho-\frac{\hbar^2}{2m}\Upsilon\frac{\D\theta}{\D
q}\;\frac{\D\rho}{\D
q}+\frac{\hbar^2}{8m}\Upsilon^2\frac{1}{\rho}\;\left(\frac
{\D\rho}{\D q}\right)^2\right\}-\\-
U\rho-\frac{\hbar^2}{8m}\;\alpha^2
\frac{1}{\rho}\left(\frac{\D\rho}{\D q}\right)^2.
\end{multline}

Varying the action~ $\EuScript{S}$ c $\widetilde{\mathcal
L}_{\scriptscriptstyle{T}} $ of form~(33) with respect to~$ \theta$ again leads to the Fokker--Planck equation similar to~ (29) but with the replacement of~$D_{qu} $ with the effective diffusion coefficient~$D_{ef}$ in it:
\begin{equation}\label{34}
\frac{\D\rho}{\D t}+\frac{\D}{\D q}\left(\rho\frac{\hbar}
{m}\frac{\D\theta}{\D q}\right)-D_{ef}\frac{\D^2\rho}{\D q^2}=0.
\end{equation}
Accordingly, varying~ $\EuScript{S}$ with respect to~$\rho$ leads to
the Hamilton--Jacobi equation generalized to the case of the
stochastic influence of the thermal vacuum:
\begin{multline}\label{35}
\hbar\frac{\D\theta}{\D t}+\frac{\hbar^2}
{2m}\left(\frac{\D\theta}{\D q}\right)^2+\frac{\hbar^2} {2m}
\Upsilon\;\frac{\D^2\theta}{\D
q^2}+U(q)-\\-\frac{\hbar^2}{8m}\Xi_T\left[\frac{1}{\rho^2}\;
\left(\frac{\D\rho}{\D q}\right)^2 +2\frac{\D}{\D
q}\left(\frac{1}{\rho}\frac{\D\rho}{\D q}\right)\right]=0,
\end{multline}
where
\begin{equation}\label{36}
\Xi_T=2 \Upsilon^2-1=2\coth^2(\varkappa \frac\omega
T)-1;\;\;\;\;\;\;\;\;\;\;\;\Xi_0=1.
\end{equation}
Obtained system of equations~ (34) and~(35) generalizes Eqs.~(22а)
and~(23а), allowing one to take the stochastic influence of the
thermal vacuum into account consistently. It is indirectly
represented in the quantities~$D_{ef}$,~$\Xi_T$, and~$\Upsilon$
contained in these equations and dependent on the world constants~
$\hbar$ and~$k_B$.

Of course, the set of Fokker--Planck equations~ (34) and
Hamilton--Jacobi equations~(35) is a nontrivial generalization
of the Schr\"odinger equations. They can be used in two ways.
This system can be solved directly for the unknown functions~$
\rho$ and~$ \theta$. We showed recently~[15] that this allows
obtaining nonequilibrium wave functions whose amplitudes and
phases are temperature-dependent and calculating
macroparameters in the nonequilibrium states using them.

However, these equations can also be modified by representing them
in the form of equations of two-velocity stochastic hydrodynamics
for the characteristic velocities~$v$ and~$u$. We see below that
these equations are a generalization of the corresponding equations
of Nelson's stochastic mechanics [10].

\section*{\small 6. One-dimensional model of two-velocity stochastic hydrodynamics}

System of equations~ (32) and~(33) can be solved directly as coupled
equations for the dissimilar variables~$\rho$ and~$\theta.$ However,
the next step can be done, and these equations can be represented in
the form of the equations for the similar variables, namely, the
velocities~$v$ and~$u_{ef}$, which are typical of any Markov
processes. In this case, we obtain the system of equations of
two-velocity stochastic hydrodynamics generalizing the equations of
Nelson's stochastic mechanics to the case of the quantum thermal
environmental influence.

We now show that Eqs.~ (34) and~(35) indeed allow obtaining the equations
of stochastic hydrodynamics in the most convenient form. Taking the foregoing
into account, we are now dealing only with one-dimensional model. To perform
the corresponding transformation, we first represent continuity equation~(34)
in the form of an equation for the diffusion velocity, which can be written
in the form
$$u_{ef}=- D_{ef}\dfrac{\D\log\rho}{\D q}$$ according to~(25).

For this purpose, we first transform Eq.~ (34) by introducing $v$
and~$u_{ef}$ explicitly in it and then multiply it by~$(-D_{ef}/\rho)$:

\begin{equation}\label{37}
-\frac{D_{ef}}{\rho}\cdot\frac{\D\rho}{\D
t}-\frac{D_{ef}}{\rho}\left[\rho\frac{\D(v+u_{ef})}{\D
q}+\frac{\D\rho}{\D q}(v+u_{ef})\right]=0.
\end{equation}
Then we differentiate the result with respect to~ $q$,
change the order of differentiation in the first term,
and transform $\log \rho,$ where it is possible, which
allows introducing $u_{ef}$ everywhere. As a result, we obtain

\begin{equation}\label{38}
\frac{\D u_{ef}}{\D t}+\frac{\D}{\D q}(vu_{ef})+\frac{\D}{\D
q}u_{ef}^2- D_{ef}\frac{\D^2}{\D q^2}(v+u_{ef})=0.
\end{equation}
We can represent this equation in a more elegant form by
forming the substantial derivarive of the diffusion
velocity~$u_{ef}$, which is typical of hydrodynamics:

\begin{equation}\label{39}
\frac{du_{ef}}{dt}\equiv\frac{\D u_{ef}}{\D t}+u_{ef}\frac{\D
u_{ef}}{\D q} =-\frac{\D}{\D q}(vu_{ef})-\frac{\D}{\D
q}\frac{u_{ef}^2}{2}+ D_{ef}\frac{\D^2}{\D q^2}(v+u_{ef}).
\end{equation}

To represent Eq.~ (35) in the explicit hydrodynamic form,
we also rewrite it in terms of the variables~$v$ and~$u_{ef}$:

\begin{equation}\label{40}
\hbar\frac{\D\theta}{\D t}+\frac m2 v^2+\frac\hbar 2
\Upsilon\frac{\D v}{\D q}+ U(q)-\frac m2\Xi_T\cdot(u_{ef}^2-\frac
\hbar m\frac{\D u_{ef}}{\D q})=0.
\end{equation}
To exclude the function~ $\theta$, we differentiate Eq.~(40)
with respect to~$q$, change the order of differentiation in the
first term, and introduce the function~$v$ explicitly into it
according to~ (24). As a result, also forming the substantial
derivative of the drift velocity~$v$ in it, we obtain
\begin{equation}\label{41}
\frac{d v}{d t} \equiv \left(\frac{\D v}{\D t}+v\frac{\D v}{\D
q}\right)=- \frac 1m\frac{\D U}{\D q}+\Xi_T\frac{\D}{\D q}
\frac{u_{ef}^2}{2}-\frac{\hbar}{2m}\left(\Xi_T\frac{\D^2 u_{ef}}{\D
q^2}+\Upsilon\frac{\D v^2}{\D q^2}\right).
\end{equation}

We recall that, as the velocity~$v$, only the quantity~$\Delta v$
generated by the stochastic action is contained in these equations.
In the case under consideration, as at~ $T=0$, the expressions
for~ $\rho$ and~$\theta$ are related to wave functions of thermal
correlated coherent states in which the exponent of the exponential
depends on~$q^2$ [5],[15]. It hence follows that the last terms in
Eqs.~(39) and~(41) containing the second derivatives of
~$v\equiv\Delta v$ and~$u_{ef}$ with respect to~ $q$ vanish.

As a result, the system of equations for the one-dimensional
model of two-velocity stochastic hydrodynamics become
\begin{equation}\label{42}
\begin{cases}
\dfrac{d u_{ef}}{dt}\equiv \dfrac{\D u_{ef}}{\D
t}+\dfrac{\D}{\D q}\dfrac{u_{ef}^2}{2} =-\dfrac{\D}{\D
q}(vu_{ef})- \dfrac{\D}{\D
q}\dfrac{u^2_{ef}}{2};\\
\dfrac{dv}{dt}=-\dfrac 1m \dfrac{\D U}{\D q}+\Xi_T\dfrac{\D}{\D q}\dfrac{u^2_{ef}}{2}.\\
\end{cases}
\end{equation}

Self-diffusion in the thermal vacuum characterized by the
coefficient~ $D_{ef}$ contained in~$u_{ef}$ is taken into account in
both these equations. In addition, the right-hand side of the lower
equation of the system contains not only the gradient of the
classical potential~ $U(q)$, but also the gradient of the energy
density of the diffusion pressure taking into account the stochastic
influence of the quantum thermostat in the cases including the
case~$T=0$. A similar contribution of the energy of the
quantum-thermostat diffusion pressure is also contained in the
right-hand side of the upper equation in~(42); in this case, it does
not vanish even at~ $v=0$.

For the comparison with the equations of Nelson's stochastic mechanics
\begin{equation}\label{43}
\left\{
\begin{array}{lcr}
\dfrac{\D u}{\D t}=-\dfrac{\D}{\D q}(vu);  \\
\dfrac{dv}{dt}=-\dfrac 1m  \dfrac{\D U}{\D q}+\dfrac{\D}{\D q}
\dfrac{u^2}{2} \\,
\end{array}
\right.
\end{equation}
we consider system of equations~ (42) in the case of the cold vacuum
($T=0$) where $u_{ef}$ transforms into~ $u$. In addition, for the
convenience of comparison, we return to the partial time derivative
in the upper equation of this system; to do this, we combine similar
terms in formula~(42). As a result, we have
\begin{equation}\label{44}
\begin{cases}
\dfrac{\D u}{\D t}=-\dfrac{\D}{\D q}(vu)-\dfrac{\D}{\D
q} u^2;\\
\dfrac{dv}{dt}=-\dfrac 1m \dfrac{\D U}{\D q}+\dfrac{\D}{\D q}
\dfrac{u^2}{2}.\\
\end{cases}
\end{equation}

As can be seen, the lower equations in systems~ (43) and~(44)
are identical completely.  However, there is an important distinction
between these systems of equations. As was expected, it is related to
the fact that self-diffusion in the cold vacuum is taken into account
in our theory; as a result of this, the equation for the diffusion
velocity contains the gradient of the energy density of the diffusion
pressure. Of course, unlike~(43), our proposed equations~(42) are
applicable at any temperature in the general case.

\section*{\small 7.   Conclusions}
Fenyes was, probably, the first who formulated the idea of using the
Lagrangian density~ $ \mathcal{L}[\rho;\theta]$ in the quantum theory~[14].
The Fokker--Planck equation containing the total velocity~$V$ of
the probability flux with the diffusion coefficient either~
$ D_{qu}$ or~$ D_T $ follows from the his proposed expression.
However, he did not introduce the generalized diffusion coefficient~$D_{ef}$.

Unlike~[14], in our approach, we consistently take into account
quantum thermal fluctuations and the energy density of the
diffusion pressure related to the stochastic environmental
action (quantum thermostat) for~$T\geqslant0$. As a final result,
we represent the Fokker--Planck and Hamilton--Jacobi equations
in the form of system of equations~ (42) for the one-dimensional
model of two-velocity stochastic hydrodynamics. The self-diffusion
coefficient in them is determined by the effective environmental
action, which is dependent on the fundamental constant~$\varkappa.$

In our opinion, following this way, one can construct full-scale
stochastic hydrodynamics taking into account not only self-diffusion
but also shear viscosity and then use it to describe NPF. To do this,
it is necessary to pass from the lower equation in~(42) for the drift
velocity to the equation that is a generalization of the Navier--Stocks
equation to the case taking self-diffusion into account.

It follows from our analysis that the self-diffusion
coefficient~$D_{ef}=\J/m$ is, probably, the most adequate
characteristic of transport phenomena, which is important
for the description of dissipative processes in NPF. At present,
it is possible to determine it experimentally by studying
diffusion of massive quarks in quark--gluon plasma obtained
in the process of heavy atom collisions.

The authors express their deep gratitude to A. G.
Zagorodnii, I. M. Mryglod, N. F. Shul'ga, I. V.
Volovich, N. M. Plakida, and Yu. P. Rybakov
as well as to the participants of their scientific
seminars for a fruitful discussion of our results.

The work was supported by the Russian Foundation
for Basic Research (project N 10-01-90408).

\newpage
\section*{\small References }

\textbf{1.} \emph{ D.T. Son and A.O. Starinets. } Viscosity,
Black Holes and Quantum Field Theory.  Ann. Rev. Nucl. Part.
Sci., 57, 95-118 (2007);
arXiv:0704.0240v2 [hep-th].\\
\textbf{2.} \emph{ T. Sch\"afer, D. Teaney.}
Nearly Perfect
Fluidity:from Cold Atomic Gases to Hot Quark-gluon Plasmas.
ArXiv: 0904. 3107 v.2 [hep-ph].\\
\textbf{3.} \emph{M\"uller M., Schmaliam J.,FritzL.} Phys.Rev.
Lett.103,025301 (2007)\\
\textbf{4.} \emph{Sukhanov A.D.  } Teor. Mat. Pyz. 2008. 154,  1,  185   \\
\textbf{5.} \emph{Sukhanov A.D., Golubjeva O.N.    }Teor. Mat. Pyz.   2009. 160, 2, 369 \\
\textbf{6.} \emph{Landau L.D., Lifshitz E.M.}  JTP, 1957. 32, 618,  \\
\textbf{7.} \emph{F\"uhrth R.} Z. Phys. 81, 143 (1933)\\
\textbf{8.} \emph{Sukhanov A.D.}  Teor. Mat. Pyz. 2004. 139, 1,129\\
\textbf{9.} \emph{Sukhanov A.D.} Teor. Mat. Pyz. 2006, 148,2, 295\\
\textbf{10.} \emph{Nelson E. } Dynamical theory of Brownian
motion. Princeton: Princ. Univ. Press, 1967.\\
\textbf{11.} \emph{ Feynman R., Leyton R., Sands M.   .} Lectures on Physics (in Russian)  9, 1967. С. 244\\
\textbf{12.}\emph{ Blokhintzev D.I.}  Principles Problems of Quantum
Mechanics (in Russian). M., Nauka, 1966, 54\\
\textbf{13.}\emph{Kolmogoroff A.N.} Math. Ann. 1931. 104, 415
 ;ibid 1933. 108,149\\
\textbf{14.} \emph{Fenyes I.} Zs. Phys.(1952) 132. C. 81  \\
\textbf{15.} \emph{ Golubjeva O.N., Sukhanov A.D.} Particles and
Nuclei (Letters) 2011, 8,1,7
  \\
\end{document}